# The impact of CAP subsidies on the productivity of cereal farms in six European countries

Luigi Biagini
*Università degli Studi della Tuscia, Italy*

Federico Antonioli

*Joint Research Centre, European Commission, Seville, Spain*
Simone Severini
*Università degli Studi della Tuscia, Italy*

**Abstract**

Total factor productivity (TFP) is a key determinant of farm development, a sector that receives substantial public support. The issue has taken on great importance today, where the conflict in Ukraine has led to repercussions on the cereal markets.

This paper investigates the effects of different subsidies on the productivity of cereal farms, accounting that farms differ according to the level of TFP. We relied on a three-step estimation strategy: i) estimation of production functions, ii) evaluation of TFP, and iii) assessment of the relationship between CAP subsidies and TFP. To overcome multiple endogeneity problems, the System-GMM estimator is adopted. The investigation embraces farms in France, Germany, Italy, Poland, Spain and the United Kingdom using the FADN samples from 2008 to 2018. Adding to previous analyses, we compare results from different countries and investigate three subsets of farms with varying levels of TFP. The outcomes confirm how CAP negatively impacts farm TFP, but the extent differs according to the type of subsidies, the six countries and, within these, among farms with different productivity groups. Therefore there is room for policy improvements in order to foster the productivity of cereal farms.

## 1. Introduction

Total factor productivity (TFP) is considered among the most critical drivers of economic growth and welfare (Basu et al., 2022). The argument gains particular strength in the agricultural sector, challenged by the increasingly competitive and globalised economy (OECD, 2014) and battered by extreme changes in the climate pattern, besides the very recent COVID19 pandemic and the war in Ukraine with their relative disruptions in labour supply and supply chain (FAO, 2022b; Global Network Against Food Crises, 2022). Indeed, while highly-productive farms may handle challenges efficiently, the opposite is true for those with low productivity levels (Olley and Pakes, 1996; Rizov, Pokrivcak and Ciaian, 2013). The European Union (EU) identifies productivity enhancement as one of the four pillars of its economic policy, guiding structural reforms, investments, and fiscal responsibility for the incoming years (EC, 2019, 2020). Remarkably, the Common Agricultural Policy (CAP) aims at the growth of EU farms' productivity (Article 39 of the EU Treaty - EC, 2007)[1], implementing around 40 billion euros per year for the EU farming sector (Massot, 2017). Therefore, investigating how the CAP affects farms' productivity remains of pivotal political and economic importance nowadays, feeding the debate on the new CAP reform that will enter into force in 2023.

Furthermore, the current Russian invasion of Ukraine shifted the equilibrium within the cereal market, with extraordinarily high prices expected to affect food security significantly, especially in low-income countries (The Economist, 2022c, 2022a, 2022b). Indeed, the European Commission also adopted a support package to help EU farmers increase their production and face such conditions (EU, 2022). This goal could be achieved in different ways, including by fostering the productivity of farms in cereal producing countries (FAO, 2022a).

The strand of literature devoted to answering how CAP leverages farms' productivity concludes with the generally negative correlation between the two (Mary, 2013; Rizov, Pokrivcak and Ciaian, 2013; Minviel and Latruffe, 2017; Khafagy and Vigani, 2022). As stated by Khafagy and Vigani (2022) in their EU-wide analysis of CAP and productivity, subsidies differ in nature, having specific aims and hence diverse impacts on TFP and farms' behaviour, partially justifying the mixed results one may find in the literature (see, for instance: Latruffe and Desjeux, 2016; Dudu and Kristkova, 2017; Garrone *et al.*, 2019).

The study attempts to fill this gap by providing a multi-country analysis regarding the impact of different CAP subsidies. Although significant differences may remain between distinct countries, the econometric strategy is applied to minimise bias in potential estimations. Before proceeding, it is essential to mention that the current analysis does not explain the reasons for differences encountered among countries. Further research and specific additional methods are needed to address this issue.

Relying on the Farm Accountancy Data Network (FADN) of France, Germany, Italy, Poland, Spain and the UK, the considered subsidies in the following analysis are: coupled direct payments (CDP), decoupled direct payments (DDP), agro-environmental scheme payments (AES), payments for less-favoured areas (LFA), and support to farm investments and other annual payments provided by the Rural Development Policies (RDP inv and RDPa Other, respectively) - (EC, 2010). The countries differ in their farms' economic, structural and environmental conditions and how the CAP is implemented.

Adding to previous analyses, we also explore whether and how CAP's effects differ according to three levels of farms' productivity – low, medium, and high. Again, this aspect appears relevant to policymakers as it would offer incipient but insightful elements for a more tailored CAP depending on specific farm characteristics.

The estimation strategy draws heavily on the literature related to the identification of the production function (Van Beveren, 2012), particularly those studies focusing on simultaneity, measurement error, and omitted variables bias (Ackerberg *et al.*, 2007), besides endogeneity drawbacks between the TFP and CAP (Mary, 2013). The Dynamic Panel GMM estimator is then applied as the efficient econometric strategy to overcome these problems with a system of instrumental variables (SYS-GMM) (Blundell and Bond, 1998; Olper *et al.*, 2014; Biagini, Antonioli and Severini, 2020).

---

[1] https://eur-lex.europa.eu/legal-content/EN/TXT/?uri=OJ:C:2007:306:TOC

Like Mary (2013), Rizov, Pokrivcak and Ciaian (2013) and Khafagy and Vigani (2022), we observed that increasing the general CAP support would have a general detrimental effect on farms' productivity, with a diversified impact depending on the nature of the subsidy as well as the country of analysis. Interestingly, results hint at different subsidy characteristics when farms' productivity levels are considered. Indeed, the negative impact of CAP subsidies is particularly relevant for low-productive farms, while such adverse effects disappear when highly-productive farms are considered. Overall, CAP lacks the ability of a productivity-booster for least-productive farms, eventually delaying their market exit but not avoiding it. On the other hand, CAP subsidies are destined for already highly-productive farms to maintain the productivity level. As a result, these less productive farms can address economic challenges less effectively than highly productive farms, thus increasing the likelihood of exiting the market.

The remainder of the paper proceeds as follows: section 2 presents a review of the relevant literature, providing the background of the analysis; section 3 describes the data used and the applied estimation strategy; section 4 shows the results, and, finally, section 5 concludes with a discussion of results obtained and providing policy implications.

## 2. Literature review

The definition of TFP can be traced back to the seminal work of Solow (1957) as the portion of production growth that cannot be attributed to the accumulation of production factors, the so-called Solow residual. Therefore, by definition, TFP is not directly observable and is potentially affected by various technological, political and socio-economic factors. On the one hand, it could be enhanced through innovation, investments, and knowledge (research); on the other hand, restrictive labour practices, excessive regulation, or an underdeveloped economic environment may hinder it (Khafagy and Vigani, 2022).

Therefore, the estimation of TFP is a challenging econometric task that agricultural economists have assessed either using indexes (see, for instance, Baldoni and Esposti, 2021), relying on a stochastic frontier framework (Minviel and Latruffe, 2017)[2], via "Control Function Estimators" (Rizov, Pokrivcak and Ciaian, 2013; Kazukauskas, Newman and Sauer, 2014; Akune and Hosoe, 2021), and "Dynamic panel estimators", although not extensively applied in agricultural economics (Mary, 2013; Garrone *et al.*, 2019).

The correlation between TFP and CAP is even more nuanced with the transition from "coupled" to "decoupled" subsidies, as confirmed by Kazukauskas, Newman and Sauer (2014). Latruffe *et al.* (2017) assess the relationship between productivity and policy subsidy levels in the dairy farms of nine countries, yielding contrasting results with policy measures influencing farmers' behaviour and, consequently, the production process (Kumbhakar and Lien, 2010). Zhu and Oude Lansink (2008) divide the measures according to their influence on price levels, farm income, investment and market participation. Therefore, different policies can have distinguished impacts on farm productivity, making assessing the role of CAP on it very challenging.

---

[2] An estensive review about TFP estimation procedures is reported in Van Beveren (2012)

Rizov et al. (2013) found that the overall CAP support generally negatively impacts TFP in almost all countries, causing allocative and technical efficiency losses because farmers can invest excessively in subsidized inputs. Likewise, Garrone *et al.* (2019) found that CAP has no well-defined effect on labour productivity. The assessment of CAP can hide significant heterogeneity because of the different types of subsidies envisaged by the CAP. For example, decoupled subsidies positively affect labour productivity, while coupled subsidies slow down labour productivity (Garrone *et al.*, 2019). This may explain why Minviel and Latruffe (2017) reported a considerable heterogeneity of results in their extensive review of analyses on the impact of CAP support on technical efficiency. This disparateness could be because of the large array of types of subsidies and because the reviewed analyses refer to different countries and periods.

To the best of our knowledge, no studies have so far compared how the different CAP subsidies provided by the CAP on TFP differ among countries and, within these, among farms experiencing different levels of TFP.

## 3. Estimation strategy and Data

The econometric strategy we apply follows the three-step technique as in Mary (2013). The first step defines and estimates production function; in the second step, the TFP is calculated, and three different levels of productivity are defined; finally, the third step devotes to assessing the relationship between CAP subsidies and farms' productivity, hence accounting also for different TFP levels among farms within each country. The following sub-sections offer more details on the three steps mentioned above.

### 3.1 Definition and estimation of the production function (step 1)

Cobb-Douglas is the most common production function used for TFP estimation. However, as with many simple formulations, there is a trade-off between simplicity and underlying assumptions. In particular, constant returns to scale are usually imposed, which can be too restrictive. We relaxed this assumption following the solution proposed by Blundell and Bond (2000) and Mary (2013).

Recently, Khafagy and Vigani (2022) use the non-nested CES production function with three inputs – land, capital and labour – arguing that Cobb-Douglas production functions are limited by the restrictive assumption in the elasticity of substitution. However, Kim, Luo and Su (2019) have demonstrated that translog or CES functions may generate multiple spurious solutions, making estimating quite challenging. Finally, it is necessary to consider that the Cobb-Douglas production function allows comparing the results with other similar studies such as Mary (2013), Rizov, Pokrivcak and Ciaian (2013) and Akune and Hosoe (2021).

In the first step, the production function for the *i*-th farm at time *t* is estimated using the Cobb-Douglas specification as follows:

$$Y_{i,t} = \Omega_{i,t} e^{\alpha_{i,t}} K_{it}^{\beta_K} N_{it}^{\beta_N} L_{it}^{\beta_L} M_{it}^{\beta_M} G_{it}^{\beta_G} e^{\epsilon_{i,t}} \quad i = 1, \dots, N; \quad t = 1, \dots, T \tag{1}$$

where: $Y_{i,t}$ is production output, $\Omega_{i,t}$ is the level of TFP and $\alpha_{i,t}$ can be interpreted as the mean production level, $K_{i,t}$ is capital stock, $N_{i,t}$ is the amount of labour, $L_{i,t}$ represents the amount of land, $M_{i,t}$ refers to materials and $G_{i,t}$ to the total amount of CAP subsidies received by the farmer[3]. The relation between production factors $(K_{i,t}, N_{i,t}, L_{i,t}, M_{i,t}, G_{i,t})$ and $Y_{i,t}$ is given by Solow residual or TFP $\Omega_{i,t}$.

The logarithmic transformation of (1) yields the following formulation with small letters to identify the logarithmic transformation of the original variables:

$$y_{i,t} = \alpha_{i,t} + \beta_k k_{i,t} + \beta_n n_{i,t} + \beta_l l_{i,t} + \beta_m m_{i,t} + \beta_g g_{i,t} + \omega_{i,t} + \epsilon_{i,t} \tag{2}$$

Due to endogeneity issues, identifying coefficients $\beta_k, \beta_n, \beta_l, \beta_m,$ and $\beta_g$ under OLS generates biased results caused by the simultaneity. The farmer chooses inputs to maximize profits, knowing $\omega_{i,t}$ that it is not directly observable. Using OLS, one indeed assumes $\varepsilon_{i,t} = \omega_{i,t} + \epsilon_{i,t}$ including TFP in the error term and assuming it i.i.d. (Marschak and Andrews, 1944; Ackerberg *et al.*, 2007). This simplification makes the estimation bias (Ackerberg *et al.*, 2007).

Mundlak, (1963) seminal work faced the issue by using both the deviations from the firm-specific mean, including time-invariant characteristics (also known as the "within" estimator) ($\eta_i$) and time intercept ($\gamma_t$). In particular, Mundlak (1963) assumes that estimated TFP also includes in the individual time-invariant characteristics ($\eta_i + \omega_i$), thus the error term in (1) becomes $\epsilon_{i,t} = \gamma_t + (\eta_i + \omega_i) + \varepsilon_{i,t}$ where $\gamma_t$ is year-specific intercept that, in our case, should account for changes occurring over the years and affecting all farms in each country, $(\eta_i + \omega_i)$ is the individual fixed intercept $\eta_i$ plus TFP ($\omega_i$), and $\varepsilon_{i,t}$ refers to serially uncorrelated measurement errors. Therefore:

$$y_{i,t} = \alpha_{i,t} + \beta_k k_{i,t} + \beta_n n_{i,t} + \beta_l l_{i,t} + \beta_m m_{i,t} + \beta_g g_{i,t} + \gamma_t + (\eta_i + \omega_i) + \varepsilon_{i,t} \tag{3}$$

---

[3] All variables are defined as in Rizov et al., (2013). More details in the appendix A.1.

Nevertheless, this formulation does not solve the issue entirely (Ackerberg, Caves and Frazer, 2015). The use of instrumental variables (IV) has emerged as one solution for the endogeneity issue, although the identification of IVs is complex and does not entirely solve the problems of endogeneity (Ackerberg, Caves and Frazer, 2015).

The shortfall in overcoming endogeneity problems using the "within" estimator or IVs is explained in Olley and Pakes (1996) and Ackerberg *et al.* (2007). $k$ and $l$ are not fully controled by the farmer a they dependend on past values (i.e. state variables):for example, in the case of $k_{i,t} = k_{i,t-1} + Investments_{i,t} - Depreciation_{i,t}$ or $k_{i,t} = \rho k_{i,t-1} + \varepsilon_{i,t}$[4]. The same argumentation applies to land. In contrast, other inputs such as $n$, $m$ and $g$ are under farmers' control (i.e. free-state variables). Because of this, state variables are less correlated with $\varepsilon_{i,t}$, engendering lower estimation bias. On the other hand, free-state variables are strongly correlated with residuals. Therefore, using a generic IV for all inputs is not the best choice as it generates biased estimates.

An additional and essential aspect to consider is the dynamic nature of the process, hence the relationship between productivity and CAP subsidies. Farmers' choices are not independent of their experience and knowledge, which require them to encompass past behaviour. To address this issue, the literature offers two main econometric strategies: the "control function estimate" (CFE) and the "dynamic panel" approach.

Olley and Pakes (1996) firstly implemented CFE, assuming TFP is a Markov process and using a two-step semiparametric procedure. However, such a procedure has been criticized for using investment as a proxy variable for TFP: first, for many points in time investment values would be zero; secondly, investments may not be an efficient IV for TFP annual changes as investing is a multi-year process to decide, improve and repay (Levinsohn and Petrin, 2003; Ackerberg *et al.*, 2007; Ackerberg, Caves and Frazer, 2015). Eventually, Levinsohn and Petrin (2003) and Ackerberg, Caves and Frazer (2015) overcome this critique by using materials as a proxy for TFP and other timing assumptions.

The "dynamic panel" approach we use relies on Arellano and Bond (1991), refined by Blundell and Bond (1998) and finally by Blundell and Bond (2000)[5]. Blundell and Bond (2000) allow estimating a production function using the assumption that $\omega_{i,t} = \rho \omega_{i,t-1} + \xi_{i,t}$ with $|\rho| < 1$, where $\xi_{i,t}$ is the productive shock. We also assume that the error terms are $\varepsilon_{i,t}, \xi_{i,t} \sim MA(0)$ and $\xi_{i,t} \sim N(0, \sigma^2)$. This allows using the following double-differencing dynamic representation:

---

[5] The use of Blundell and Bond, (1998) estimators requires to verify the presence of unit root as also highlited by Mary, (2013). The unit-root test is conducted as per Bond and Söderbom, (2005). Results are available upon request.

$$y_{i,t} = \pi_1 y_{i,t-1} + \pi_2 k_{i,t} + \pi_3 k_{i,t-1} + \pi_4 l_{i,t} + \pi_5 l_{i,t-1} + \pi_6 n_{i,t} + \pi_7 n_{i,t-1} + \pi_8 m_{i,t}$$
$$+ \pi_9 m_{i,t-1} + \pi_{10} g_{i,t} + \pi_{11} g_{i,t-1} + (1-\rho)\eta_i + (\omega_{i,t} - \rho\omega_{i,t-1}) \quad (4)$$
$$+ (\varepsilon_{i,t} - \rho\varepsilon_{i,t-1}) + \xi_{i,t}$$

This model addresses simultaneity and omitted variables bias by using both lagged levels and first differences as IVs. Recently, Roodman (2009a, 2009b) and Windmeijer (2005, 2021) have updated this estimation, proposing a better instrumenting strategy that makes the estimated errors more robust. The battery of tests described in Arellano and Bond (1991) allows checking for the robustness of estimation, in particular for autoregressive component (m1 or AR1 – m2 or AR2) and the correct specification of IV (i.e., the Sargan test). While CFE requires the researcher to identify the free-, state- and proxy- factors, dynamic panel estimators do not (Ackerberg, 2017, 2020), with the risk of generating evaluation errors. For instance, labour inputs are commonly assumed to be a free variable, while part of this is determined by the past, such as the amount of family unpaid work, which is particularly important in agriculture. Therefore, such an assumption concerning one of the production factors may lead to biased estimates. Finally, dynamic panel estimators reduce errors from omitted variables and measurement errors via the Windmeijer correction, using the Generalized Method of moments (GMM) to avoid specification errors in the functional form and a system of IVs, with two-step procedure and Windmeijer correction to reduce the problems of omission of variables and of measurement on the input side (Lizal and Galuscak, 2012; Song, 2015; Kim, Petrin and Song, 2016).

## 3.2 Estimation of TFP (Step 2)

For measuring farm-level productivity, we rely on the Solow Residual (Olley and Pakes, 1996; Ackerberg *et al.*, 2007). From equation (4) $\pi_1 = \rho$, $\pi_2 = \beta_k$, $\pi_3 = \rho\beta_k$, $\pi_4 = \beta_l$, $\pi_5 = \rho\beta_l$, $\pi_6 = \beta_n$, $\pi_7 = \rho\beta_n$, $\pi_8 = \beta_m$, $\pi_9 = \rho\beta_m$, $\pi_{10} = \beta_g$, $\pi_{11} = \rho\beta_g$. To estimate $\hat{\beta}_k, \hat{\beta}_l, \widehat{\beta}_n, \widehat{\beta}_m, \widehat{\beta}_g$ the following minimum distance estimator (Blundell et al., 1996; Chamberlain, 1982) (see appendix A.2):

$$TFP_{i,t} = exp\left(y_{i,t} - \hat{\beta}_k k_{i,t} + \hat{\beta}_l l_{i,t} + \widehat{\beta}_n n_{i,t} + \widehat{\beta}_m m_{i,t} + \widehat{\beta}_g g_{i,t}\right) \quad (5)$$

## 3.3 Assessing the relationship between TFP and different CAP measures (Step 3)

The linkage between CAP and its specific measures and the farms' productivity is disentangled within the third and last methodological step. Relying once more on the dynamic panel estimator, the TFP estimated in Step 2 is represented as:

$$TFP_{i,t} = f\big(\text{Economic Dimension}_{i,t}, CDP_{i,t}, DDP_{i,t}, AES_{i,t}, LFA_{i,t}, RDPa\ others_{i,t}, RDP\ inv_{i,t}\big) \quad (6)$$

As in Mary (2013), the GMM-SYS estimator accounts for many factors that are not directly observable (e.g., location of the farm in marginal zones, the extent of mechanization, or soil quality)

Many farm-specific factors (for instance, whether located in marginal areas, the degree of mechanization, or soil quality) are also accounted for GMM-SYS estimator.

The different nature of CAP subsidies implies that the subsidies are endogenous variables (Mary, 2013). For instance, subsidies coupled to production volumes have been assessed based on regional-yield levels; aids for less-favoured areas require the farm to be located in a specific geographical location to receive the subsidy; some incentives belonging to Pillar 2, such as those intended to cover farmers' investments, require the economic ability to anticipate the investment. Therefore, neglecting the presence of such endogenous processes would lead to inaccurate estimates of the relationship. Again, the GMM-SYS estimator offers a solution by instrumenting endogenous variables and considering some CAP subsidies' multi-year (dynamic) nature (Mary, 2013).

The analysis is developed first on each country's total sample of farms. Nevertheless, farm-specific productivity conditions may shape the relationship between the TFP and the CAP differently, and subgroups of low, medium, and high productivity. These subgroups are defined using the TFP estimated in Step 2 (Solow residual) after the production function estimation via the CFE approach by Ackerberg *et al.* (2015) ($TFP_{ACF}$)[6]. The use of this version of TFP looks pretty appropriate for segmenting farms because it provides the overall level of TFP that includes the fixed component that, in contrast, is cleaned out when using the panel dynamic estimator in Step 3 (Lee, Stoyanov and Zubanov, 2019; Abito, 2020; Ackerberg, 2021). Considering that the $TFP_{ACF_{i,t}}$ can vary between years, the median value of each farm has been used for the categorization of the farm : $TFP_{ACF}^{Low}, TFP_{ACF}^{Medium}, TFP_{ACF}^{High}$. Accordingly, the TFP is now a function of CAP measures and conditionally on three different productivity-related groups (7):

$$TFP_{i,t} = f\big(\text{Economic Dimension}_{i,t}, CDP_{i,t}, DDP_{i,t}, AES_{i,t}, LFA_{i,t}, RDPa\ others_{i,t}, RDP\ inv_{i,t}\big) | TFP_{ACF}\ group \quad (7)$$

It is essential to underline that these three levels are defined within each country separately. Therefore, the average level of one class of TFP (e.g., low) in one country may differ from that of the farms classified in the same group in another country. This should be bear in mind when interpreting the results.

### 3.4 Data

The Farm Accountancy Data Network (FADN) for 2008 – 2018 serves as the basis for the analysis.

---

[6] See Appendix for more details.

While some policy changes occurred during the period, the inclusion of yearly dummies should account for such changes that did not completely change the nature of the different measures we considered.

Specialised cereal farms are investigated. This ensures homogeneity in both technology of production and output, particularly relevant when estimating production function (Mary, 2013; Rizov, Pokrivcak and Ciaian, 2013; Latruffe *et al.*, 2017). Using all FADN observations would not allow correctly estimating TFP because of omitted variables (prices) bias. Both Mary (2013) and Rizov *et al.* (2013) do not estimate TFP using production levels, but revenues inclusive of the price of inputs ($k$, $l$ and $m$) as quantities data are not available. In order to reduce the distortions of omitted prices bias, we use the specific price indices. (Zhu and Oude Lansink, 2008; Loecker, 2011; Rizov, Pokrivcak and Ciaian, 2013; Bond *et al.*, 2021).

The set of farms selected for the analysis has been chosen to be as homogeneous as possible as well as representative of the cereal sector to grasp interesting insights and draw significant policy-relevant conclusions.[7] Finally, the econometric strategy conditions the selection of the analysed countries: a sufficiently large number of observations over time is required, so field crop farms in France, Germany, Italy, Poland, Spain and the UK were analysed. Indeed, the number of observations for these countries ranges from 19,248 in France to 31,819 in Polland. In contrast, a lower number of observations is available for the UK. This suggests that the sample, in most cases, is large enough to represent an essential share of the cereals farms of each country[8].

According to Eurostat, averaging on 2008-2018, these six countries together account for around 65% and 67% of EU cereal crops' area and harvested production of the whole EU, respectively, representing a highly significant group of producing countries[9] (Figure 1).

**Figure 1 - Cereal production in some European countries and the EU, 2008-2018.**

---

[7] General type of farming '1. Specialist field crops', principal type of farming '15. Specialist cereals, oilseeds and protein crops' and '16. General field cropping' according to the Commission Implementing Regulation (Eu) 2015/220 (available at: https://eur-lex.europa.eu/legal-content/EN/TXT/PDF/?uri=CELEX:32015R0220&from=EN) (see Annex XV).

[8] This does not necessarily imply that the selected samples perfectly represent the whole population of the cereal farms of each country.

[9] Crop production in EU standard humidity [apro_cpsh1] (available at: https://ec.europa.eu/eurostat/databrowser/view/apro_cpsh1/default/table?lang=en).

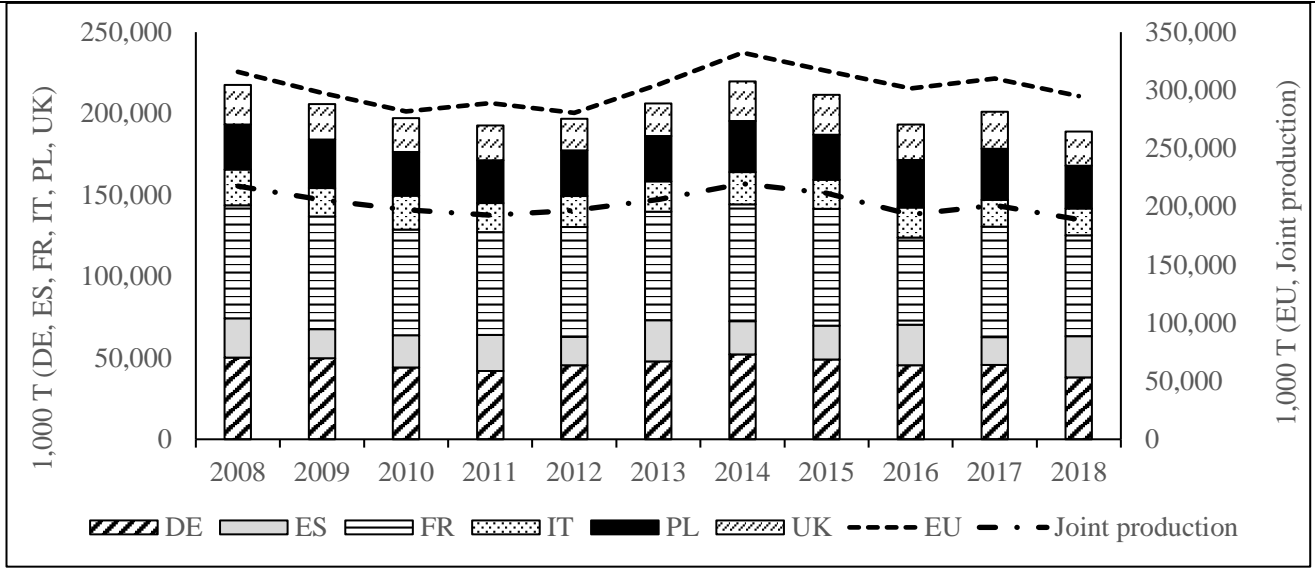

*Source: authors' elaboration on Eurostat data.*

The farms in these countries differ in economic and structural characteristics (Table 1, Table A. 2, and Table A. 3). On average, farms in the UK are larger than in other countries, while the opposite is true for those in Spain. Farms in the UK also use a relatively more limited amount of labour than in other countries. Furthermore, France's output per unit of land is very high compared to other countries. Finally, the relative importance of CAP is similar, on average, in the farms located in the six countries, with the UK showing a slightly lower relative importance of CAP subsidies (i.e. in comparison with the generated total output).

Furthermore, the six countries also differ in how the CAP has been implemented. Regarding the basic payment, for example, the UK opted for a regional flat-rate payment, Poland supported farmers through the Single Area Payment Scheme, while the other countries opted for a partial convergence approach (Henke *et al.*, 2015). The six countries also differ regarding the relative importance of coupled direct payments. These represent a relatively high share of the overall financial ceiling devoted to direct support in France and, to a lower extent, Poland, Spain and Italy, while a way lower share in UK and Germany (Henke *et al.*, 2015).

*Working Paper*

**Table 1- General Statistics**

| | UM | Germany (DE) | | Spain (ES) | | France (FR) | | Italy (IT) | | Poland (PL) | | United Kingdom (UK) | |
|---|---|---|---|---|---|---|---|---|---|---|---|---|---|
| **NR. OBS.** | Number | 22,390 | | 20,528 | | 19,248 | | 29,608 | | 31,819 | | 5,261 | |
| **TOTAL OUTPUT** | Euro | 331,579 | *(636,082)* | 66,000 | *(100,771)* | 220,617 | *(202,512)* | 95,835 | *(255,072)* | 88,487 | *(222,516)* | 471,643 | *(717,780)* |
| **CAPITAL** | Euro | 1,045,738 | *(1,365,109)* | 267,634 | *(387,432)* | 264,264 | *(239,343)* | 571,425 | *(2,011,861)* | 364,251 | *(498,364)* | 311,2078 | *(3,981,256)* |
| **LAND** | Euro | 341,163 | *(725,325)* | 146,172 | *(329,149)* | 42,823 | *(93,252)* | 336,224 | *(1,158,662)* | 122,850 | *(303,510)* | 1,477,048 | *(2,678,486)* |
| **LABOUR** | Hours | 6,092 | *(12,699)* | 2,608 | *(2,936)* | 3,037 | *(3,096)* | 3,548 | *(5,054)* | 5,263 | *(8,495)* | 7,084 | *(13,863)* |
| **MATERIAL** | Euro | 205,906 | *(387,037)* | 39,907 | *(56,301)* | 137,494 | *(108,411)* | 46,164 | *(115,817)* | 55,537 | *(162,092)* | 287,600 | *(390,341)* |
| **CAP SUBSIDIES** | Euro | 72,181 | *(129,541)* | 20,904 | *(26,606)* | 44,814 | *(30,656)* | 23,037 | *(52,062)* | 23,313 | *(46,427)* | 80,060 | *(96,136)* |

*Source:* authors' *elaboration on FADN data* - *Note: Standard deviation in brackets*

-11-

## 4. Estimation results

The results of production function estimates (Step 1) are reported in Table 2; Table 3 details the SYS-GMM TFP estimation (Step 2); Table 4 and Table 5 feature the impacts of CAP measures on the farms' TFP (Step 3) for each country and the whole sample but according to different TFP levels, respectively.

### 4.1 Production function estimates (Step 1)

Looking at Table 2, the coefficient $y_{t-1}$ is statistically significant for Spain, Poland and Italy and Germany, signalling that a share of the total output persists through time, hence explaining the level of the total outcome of the following period ($y_t$). For instance, between 20% and 30% of production output ripened the previous year is transferred to the output of the following campaign, hinting at certain resilience to shocks.

All model specifications successfully fulfil the usual battery of statistical tests: the autoregressive error component tests of orders 1 and 2 (AR1 and AR2) have the expected significance levels – i.e., highly significant for AR (1), thus ascertaining the presence of an autoregressive process of order one, and not significant for AR (2), excluding the autoregressive process of order more than one – as well as Arellano and Bond (1991), Blundell and Bond (1998) and Roodman (2009a, 2009b). The (non)significance of the Sargan test suggests the right set of instruments for correcting for the (potential) endogeneity has been used in the model estimation, passing the overidentifying restrictions test hence failing to reject the null of valid overidentifying restrictions. Wald Test on coefficients demonstrates that models' specifications differ by using the only constant term, while the Wald Test on time dummies highlights the existence of year-specific effects (Arellano and Bond, 1991; Blundell and Bond, 1998; Roodman, 2009b, 2009a).

### 4.2 Total factor productivity estimates (Step 2)

Looking at Table 3, results from the TFP estimation underline the highest values for Germany and Spain, followed by Italy and Poland and, finally, France, whose results are in line with Mary (2013) and the UK. Figure 2 depicts the TFP trend for each country, interestingly showing quite similar pathways. 2014 seems to represent a unique and common shock for all countries, with all having restored to their previous levels of TFP at the end of the period analysed, demonstrating the sectorial resilience, thus the capacity to revert to its initial state. According to Ackerberg et al. (2007), when differences arise between TFP levels at different points in time, these are due to farm innovations, for either positive or negative differences - for instance, when the farmer introduces a new technology or crop diversification, this may negatively affect productivity.

**Table 2 - Production function estimation using Blundell – Bond (2000) Approach – (Step 1)**

|  |  | Germany (DE) |  |  | Spain (ES) |  |  | France (FR) |  |  | Italy (IT) |  |  | Poland (PL) |  |  | United Kingdom (UK) |  |  |
|---|---|---|---|---|---|---|---|---|---|---|---|---|---|---|---|---|---|---|---|
| $y_{(t-1)}$ |  | 0.278 | *(0.080)* | *** | 0.189 | *(0.029)* | *** | 0.057 | *(0.089)* |  | 0.206 | *(0.033)* | *** | 0.117 | *(0.050)* | * | -0.054 | *(0.102)* |  |
| $k_{(t)}$ |  | -0.596 | *(0.401)* |  | -0.648 | *(0.566)* |  | -0.206 | *(0.181)* |  | -0.003 | *(0.120)* |  | -1.124 | *(0.442)* | * | 0.038 | *(0.076)* |  |
| $k_{(t-1)}$ |  | 0.614 | *(0.368)* |  | 0.512 | *(0.530)* |  | 0.161 | *(0.155)* |  | 0.036 | *(0.106)* |  | 1.047 | *(0.400)* | ** | 0.017 | *(0.048)* |  |
| $l_{(t)}$ |  | 0.084 | *(0.137)* |  | -0.085 | *(0.209)* |  | 0.186 | *(0.164)* |  | 0.096 | *(0.084)* |  | -0.102 | *(0.121)* |  | 0.002 | *(0.073)* |  |
| $l_{(t-1)}$ |  | -0.113 | *(0.128)* |  | 0.033 | *(0.193)* |  | -0.189 | *(0.151)* |  | -0.090 | *(0.076)* |  | 0.069 | *(0.110)* |  | 0.024 | *(0.070)* |  |
| $n_{(t)}$ |  | -0.497 | *(0.286)* |  | 0.569 | *(0.366)* |  | 1.267 | *(0.601)* | * | 0.764 | *(0.345)* | * | -0.043 | *(0.563)* |  | 0.620 | *(0.465)* |  |
| $n_{(t-1)}$ |  | 0.604 | *(0.219)* | ** | -0.062 | *(0.103)* |  | -0.914 | *(0.479)* |  | -0.273 | *(0.143)* |  | 0.159 | *(0.359)* |  | -0.336 | *(0.331)* |  |
| $m_{(t)}$ |  | -0.185 | *(0.471)* |  | 0.615 | *(0.126)* | *** | 0.715 | *(0.395)* |  | 0.730 | *(0.140)* | *** | 0.855 | *(0.263)* | ** | 1.335 | *(0.481)* | ** |
| $m_{(t-1)}$ |  | 0.398 | *(0.150)* | ** | -0.054 | *(0.055)* |  | -0.067 | *(0.146)* |  | -0.140 | *(0.052)* | ** | -0.058 | *(0.091)* |  | -0.272 | *(0.183)* |  |
| $g_{(t)}$ |  | 0.480 | *(0.325)* |  | -0.051 | *(0.092)* |  | 0.387 | *(0.215)* |  | -0.046 | *(0.121)* |  | 0.275 | *(0.156)* |  | 0.005 | *(0.617)* |  |
| $g_{(t-1)}$ |  | -0.068 | *(0.087)* |  | 0.058 | *(0.035)* |  | -0.037 | *(0.071)* |  | 0.024 | *(0.052)* |  | 0.037 | *(0.059)* |  | -0.100 | *(0.279)* |  |
| Sargan Test | p-value | 0.195 | | | 0.324 | | | 0.893 | | | 0.241 | | | 0.208 | | | 0.818 | | |
| AR (1) | p-value | 0.001 | | | 0.000 | | | 0.004 | | | 0.000 | | | 0.000 | | | 0.001 | | |
| AR (2) | p-value | 0.930 | | | 0.375 | | | 0.080 | | | 0.909 | | | 0.434 | | | 0.471 | | |
| Wald (Coefficients) | p-value | 0.000 | | | 0.000 | | | 0.000 | | | 0.000 | | | 0.000 | | | 0.000 | | |
| Wald (Time Dummies) | p-value | 0.000 | | | 0.000 | | | 0.000 | | | 0.000 | | | 0.000 | | | 0.000 | | |

*Notes: \*\*\* $p < 0.001$; \*\* $p < 0.01$; \* $p < 0.05$ - Standard errors in brackets in italics*   *Source: our elaboration on FADN data*

**Table 3 - TFP Estimation using Blundell Bond (2000) Approach**

| Year | GERMANY (DE) | SPAIN (ES) | FRANCE (FR) | ITALY (IT) | POLAND (PL) | GERMANY (DE) |
|---|---|---|---|---|---|---|
| 2009 | 1.549 | 1.422 | 0.878 | 1.068 | 1.053 | 0.807 |
| 2010 | 1.592 | 1.433 | 0.903 | 1.066 | 1.018 | 0.823 |
| 2011 | 1.589 | 1.432 | 0.907 | 1.065 | 1.005 | 0.826 |
| 2012 | 1.611 | 1.424 | 0.912 | 1.063 | 1.014 | 0.819 |
| 2013 | 1.599 | 1.422 | 0.898 | 1.060 | 1.006 | 0.816 |
| 2014 | 1.683 | 1.363 | 0.917 | 1.102 | 0.964 | 0.811 |
| 2015 | 1.556 | 1.387 | 0.901 | 1.069 | 0.993 | 0.819 |
| 2016 | 1.557 | 1.403 | 0.889 | 1.069 | 0.979 | 0.820 |
| 2017 | 1.554 | 1.394 | 0.900 | 1.066 | 0.991 | 0.820 |
| 2018 | 1.550 | 1.405 | 0.902 | 1.068 | 0.985 | 0.825 |

*Source: our elaboration on FADN data*

**Figure 2 - TFP estimates by country in the period 2008-2018.**

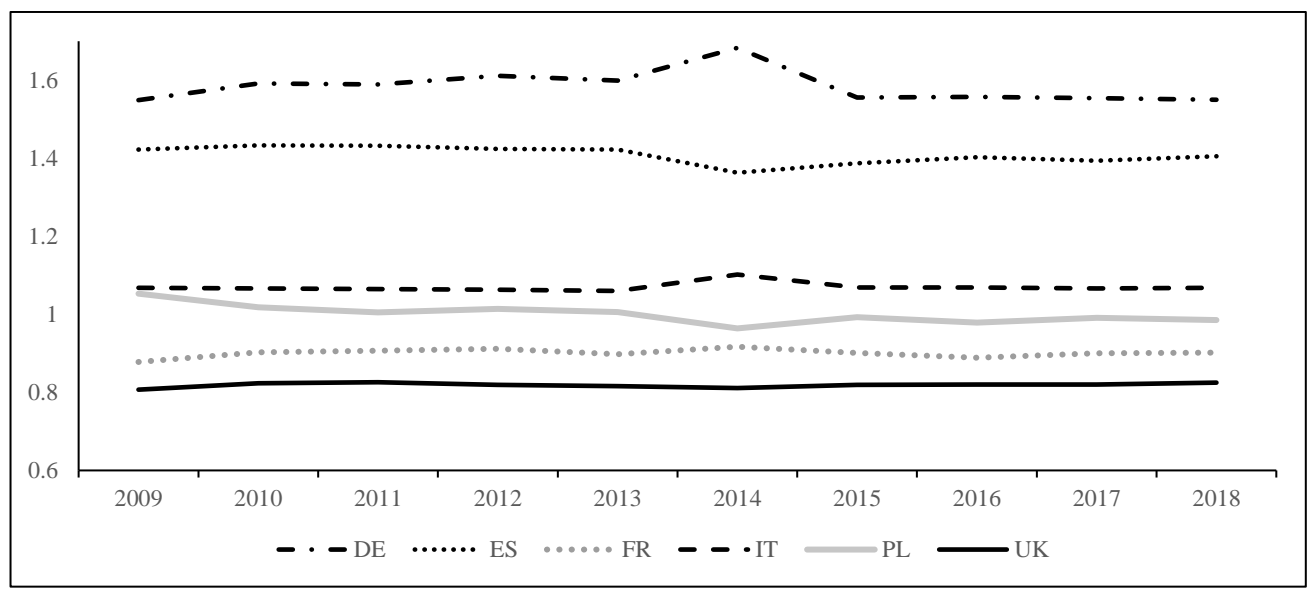

*Source: our elaboration on FADN data*

## 4.3 The impact(s) of CAP subsidies on TFP (Step 3)

The results of the third step of the empirical application show the impact of the different CAP payments on the TFP per country and, within each, per different TFP levels as defined in section 3.3. Results are synthesized graphically in Table 5, while the more detailed results are provided in the Tables in Appendix A.3. Results show that the lagged term of TFP is not significant for most of the productivity categories. Indeed, it turns significant only when the $TFP_{ACF}^{High}$ group is concerned, being positive for Spain, Italy, Poland and the UK. In contrast, this is negative for France and not significant for Germany. This may suggest that farms within the low – medium range have not yet reached a stable productivity level, oscillating from year to year, while the opposite is true for the most productive agricultural units.

Economic sizes - recalling that this is a continuous measure expressed by the standard output in thousands of euros, when statistically significant, shows a positive relationship with the TFP for any class of productivity. Interestingly, this is often true for farms belonging to the lower segment of TFP and not necessarily those belonging to the higher segment, at least in Germany and Italy. As expected, the larger the farms in terms of output, the greater its positive leverage on TFP, mainly when the unit operates at a low level of TFP. For farms that already rely on high productivity levels, the economic size effect can still be significant and positive, as in Spain and France. Peculiar is the result for the UK, where the economic size points to a negative effect on TFP for both medium- and high-productivity levels.

Looking at the different CAP subsidies, those belonging to Pillar 1 of the CAP, CDP and DDP bear negligible or negative effects on the TFP for all countries analyzed and all productivity levels. Germany, Italy, the UK, and Spain feature no effects except a mild negative influence on medium-TFP Spanish farmers and medium and high-TFP French farms. Being CDP subsidies coupled to the quantity produced, while it is in line with expectations, a non-significant effect (farms that already exhausted their capacity to reach the productivity by definition are not able to increase continuously), negative impacts are yet controversial results; nevertheless, the recent work of Khafagy and Vigani (2022) also finds some general discouraging impacts of CDP on EU-27 farms' productivity, particularly for all but Northern EU farms. Besides, both Mary (2013) and Rizov et al. (2013) also find a negative impact of subsidies on productivity. The authors agree on the ability of subsidies to buffer negative income situations while (negatively) shaping farmers' willingness to improve their productivity. Likewise, DDP results align with the main conclusion one can draw from the literature, thus negatively influencing the farms' TFP levels. Indeed, since these payments are decoupled from the quantity produced, they are a sure asset for all cereal farmers, potentially weakening their willingness to invest to reach higher productivity levels. Both the UK and Spain do not register any significant effect for DDP. At the same time, for countries such as Germany, France, Italy, and Poland, at least one category of farms indicates a significant and negative effect.

Regarding Pillar 2 payments, the situation mirrors almost entirely that of Pillar 1, although some differences emerge. Agro-Environmental Schemes (AES) are generally negative – again Spain and the UK do not show any significant effect for this specific measure. Such payments are linked to environmentally friendly activities, hence not necessarily the most productive methods (Vigani, Rodriguez Cerezo and Gomez Barbero, 2015), and designed to provide financial support to farmers who chose to reduce negative externalities, producing public goods, despite lower productivity with respect to their conventional counterparts. Such results align with Mennig and Sauer (2020) for dairy farms, suggesting that participation in AES schemes may reduce the farm's flexibility for further development. All in all, farms satisfying the requirements for receiving the AES may be involved in agricultural tasks that are not as productive as the conventionals', explaining such a negative global effect.

Similarly, Less Favoured Areas payments (LFA) support farms located in disadvantaged areas, often starting from lower productivity performances than their counterparts (Vigani and Dwyer, 2020) as per lacking those services otherwise present in common areas. Unlike Germany, Italy, and the UK, Spain, France, and Poland appear to be negatively impacted by such subsidies. On the one hand, the absence of any effect may hint that LFA payments are supporting farms by avoiding their exit from the market, avoiding (heavier) productivity losses that endanger their survival. On the other hand, adverse effects may suggest that, depending on the area, such financial support is insufficient to stop the negative trend in productivity. Likewise, farms already operating at economic losses may not engage in any productivity-enhancing activity, receiving subsidies while waiting to be exiting the market. Again, the literature offers controversial results on these types of payments: on the one hand, our results are in line with Mary (2013) and Khafagy and Vigani (2022), but opposite to those in Dudu and Kristkova (2017) and Garrone et al. (2019).

Finally, Pillar 2 support for investments also entails a negligible-to-negative effect on productivity - negative for Germany, France, and Italy, not significant for Poland and the UK - while positive only for Spain. While one may expect positive-to-non-significant results from investment support – investments usually keep up or enhance farms' productivity – to note the misalignment between the provision of such lump sum and the investment. In other words, it may take some time for the farm to operate the investment and the latter to start fruitful. Hence, such a negative effect may be due to the initiating phase of the new investment, which may involve a change in the production system, the technology or the machinery. On the other hand, Spain's positive and significant result points to the opposite. This could be related to how the support is granted, that is, 'in advance' of the investment or when the investment is already in place. Nevertheless, results from the relevant literature found the same relationship (Lakner, 2009; Latruffe and Desjeux, 2016; Latruffe *et al.*, 2016; Nilsson, 2017; Khafagy and Vigani, 2022).

**Table 4 - Impacts of farm subsidies on TFP for six countries – Total Sample**

|  | Germany (DE) | Spain (ES) | France (FR) | Italy (IT) | Poland (PL) | United Kingdom () |
|---|---|---|---|---|---|---|
| $TFP_{t-1}$ | 0.002*** | −0.019 | −0.011 | 0.050* | 0.031 | 0.055 |
|  | (0.000) | (0.016) | (0.019) | (0.026) | (0.022) | (0.064) |
| Economic Size (x1000) | 0.289 | 0.921 | 0.212*** | 0.113 | 0.285** | 0.029 |
|  | (0.217) | (0.642) | (0.031) | (0.123) | (0.143) | (0.019) |
| CDP (x1000) | 0.004 | −0.010 | −0.002*** | −0.002 | −0.004*** | −0.001 |
|  | (0.009) | (0.010) | (0.000) | (0.002) | (0.001) | (0.001) |
| DDP (x1000) | −0.002* | 0.003* | −0.001*** | 0.001 | −0.0002 | −0.0002* |
|  | (0.001) | (0.002) | (0.000) | (0.001) | (0.001) | (0.000) |
| AES (x1000) | −0.003 | −0.024** | −0.001*** | −0.0001 | −0.003 | −0.00003 |
|  | (0.004) | (0.012) | (0.000) | (0.001) | (0.002) | (0.000) |
| LFA (x1000) | −0.013 | 0.039 | −0.003*** | −0.011 | −0.014*** |  |
|  | (0.014) | (0.030) | (0.001) | (0.008) | (0.005) |  |
| RDP Other (x1000) | −0.334 | −0.068 | −0.001*** | −0.001 | 0.003 | 0.004 |
|  | (0.641) | (0.078) | (0.000) | (0.004) | (0.003) | (0.004) |
| RDP inv (x1000) | −0.337 | −0.166 | −0.003*** | 0.006 | −0.001 | 0.005 |
|  | (0.650) | (0.110) | (0.000) | (0.007) | (0.004) | (0.004) |
| Sargan test (p.value) | 0.101 | 0.299 | 0.423 | 0.415 | 0.158 | 0.490 |
| m1 (p.value) | 0.000 | 0.027 | 0.000 | 0.000 | 0.000 | 0.000 |
| m2 (p.value) | 0.484 | 0.418 | 0.357 | 0.785 | 0.867 | 0.106 |
| Wald coeff. (p.value) | 0.000 | 0.001 | 0.000 | 0.001 | 0.000 | 0.139 |
| Wald time (p.value) | 0.000 | 0.000 | 0.000 | 0.000 | 0.000 | 0.000 |
| Observations | 3,304 | 2,948 | 2,555 | 5,548 | 5,021 | 779 |

Note *p<0.1; **p<0.05; *** p<0.01

*Source: our elaboration on FADN data*

**Table 5 - Sinthesis of the results by country. Overall sample (Total) and by TFP level (Low, Medium and High)**

| | ↑ Positive | ⇔ Non significant | ↓ Negative |
|---|---|---|---|

| | Germany (DE) | | | Spain (ES) | | | France (FR) | | | Italy (IT) | | | Poland (PL) | | | United Kingdom (UK) | | |
|---|---|---|---|---|---|---|---|---|---|---|---|---|---|---|---|---|---|---|
| | Total | | | Total | | | Total | | | Total | | | Total | | | Total | | |
| | L | M | H | L | M | H | L | M | H | L | M | H | L | M | H | L | M | H |
| TFP$_{t-1}$ | | ↑ | | | ⇔ | | | ⇔ | | | ↑ | | | ⇔ | | | ⇔ | |
| | ⇔ | ⇔ | ⇔ | ⇔ | ⇔ | ↑ | ⇔ | ⇔ | ↓ | ⇔ | ⇔ | ↑ | ⇔ | ⇔ | ↑ | ⇔ | ⇔ | ↑ |
| Economic Size | | ⇔ | | | ⇔ | | | ↑ | | | ⇔ | | | ↑ | | | ⇔ | |
| | ↑ | ↑ | ⇔ | ↑ | ↑ | ↑ | ⇔ | ↑ | ↑ | ↑ | ⇔ | ⇔ | ↑ | ⇔ | ↑ | ⇔ | ↓ | ↓ |
| CDP | | ⇔ | | | ⇔ | | | ↓ | | | ⇔ | | | ↓ | | | ⇔ | |
| | ⇔ | ⇔ | ⇔ | ⇔ | ↓ | ⇔ | ⇔ | ↓ | ↓ | ⇔ | ⇔ | ⇔ | ⇔ | ⇔ | ⇔ | ⇔ | ⇔ | ⇔ |
| DDP | | ↓ | | | ↑ | | | ↓ | | | ⇔ | | | ⇔ | | | ↓ | |
| | ↓ | ↓ | ⇔ | ⇔ | ⇔ | ⇔ | ↓ | ↓ | ⇔ | ↓ | ⇔ | ↓ | ↓ | ⇔ | ↓ | ⇔ | ⇔ | ⇔ |
| AES | | ⇔ | | | ↓ | | | ↓ | | | ⇔ | | | ⇔ | | | ⇔ | |
| | ↓ | ↓ | ⇔ | ⇔ | ⇔ | ⇔ | ⇔ | ↓ | ↓ | ⇔ | ⇔ | ⇔ | ⇔ | ⇔ | ↓ | ⇔ | ⇔ | ⇔ |
| LFA | | ⇔ | | | ⇔ | | | ↓ | | | ⇔ | | | ↓ | | | Not Available | | |
| | ⇔ | ⇔ | ⇔ | ↓ | ↓ | ↓ | ⇔ | ↓ | ↓ | ⇔ | ⇔ | ⇔ | ⇔ | ↓ | ⇔ | | | |
| RDP Other | | ⇔ | | | ⇔ | | | ↓ | | | ⇔ | | | ⇔ | | | ⇔ | |
| | ⇔ | ↓ | ⇔ | ⇔ | ↑ | ⇔ | ⇔ | ⇔ | ↓ | ↓ | ⇔ | ↓ | ⇔ | ⇔ | ⇔ | ⇔ | ⇔ | ⇔ |
| RDP inv | | ⇔ | | | ⇔ | | | ↓ | | | ⇔ | | | ⇔ | | | ⇔ | |
| | ⇔ | ↓ | ⇔ | ⇔ | ↑ | ↑ | ⇔ | ↓ | ↓ | ↓ | ⇔ | ↓ | ⇔ | ⇔ | ⇔ | ⇔ | ⇔ | ⇔ |

*Source: our elaboration on FADN data.*  *Note: Detailed results are provided in the Appendix.*

## 5. Summary and conclusions

Relying on the comprehensive and robust, but at the same time controversial results, literature on the relationship between CAP and productivity, we used micro-level data for modelling the total factor productivity of cereal farms for major European producers. A three-step procedure has been used to assess the link between CAP subsidies and total factor productivity for cereal farms in six different European countries for the 2008 – 2018 period. A Cobb-Douglas production function was estimated considering land, capital, labour, materials and subsidies of the CAP as production factors following Mary (2013) estimation strategy, and then obtaining TFP through the Solow residual and, finally, to evaluate the relationship between TFP and CAP measures through the SYS-GMM approach.

The added value of our research is first to provide a multi-country analysis of a specific type of farming. Indeed, as stated by Khafagy and Vigani (2022) in their EU-wide analysis of CAP and productivity, subsidies differ in nature, having specific aims and hence diverse impacts on TFP and farms' behaviour. Therefore, studying a homogeneous portion of farms in terms of typology may ensure higher precision and trustful estimated impacts. Although significant differences may remain between different countries, the econometric strategy applied, especially via time and individual fixed effects, account for such differences hence minimizing the bias of the estimates. Moreover, the results from the statistical tests performed after the SYS-GMM model estimation confirm our results' unbiasedness, overcoming endogeneity issues.

When looking at the impact of the diverse CAP measures on farms' productivity per country, results are in line with many of the empirical applications; both coupled and decoupled direct payments – i.e., CDP and DDP – in almost all cases depress productivity or bear no effects. As suggested by *Garrone et al.* (2019), CDP potentially disrupts the efficient allocation of inputs and outputs, and losing economic and financial constraints may also weaken the incentive of farmers to invest for higher productivity performances (Rizov, Pokrivcak and Ciaian, 2013). However, we did not find this effect in three of the six countries. On the other hand, DDP should enhance farms' capital and access to credit, boosting productivity (Kazukauskas, Newman and Sauer, 2014). Nevertheless, their tight link with land may convey their benefits to landowners instead of farmers via their capitalization into land rents (O'Neill and Hanrahan, 2016; Varacca *et al.*, 2021). Results of our analysis suggest that DDP indeed negatively affects productivity in most cases apart from Spain.

The same negligible-to-negative effect on productivity applies to all remaining CAP subsidies, namely LFA, AES, RDP investment and RDP other. While, theoretically, subsidies on investments should boost productivity, our results are consistent with the most recent literature (Lakner, 2009; Latruffe *et al.*, 2016; Nilsson, 2017). Such subsidies may push farmers to (over) invest in less productive activities, leaving aside more productive ones (Khafagy and Vigani, 2022).

The same situation relates to AES payments, which is a reasonable result as farms receiving such subsidies are focused on providing public goods (e.g., landscape, biodiversity, resources' quality), and it is more restricted in their management development (Mennig and Sauer, 2020). Finally, LFA subsidies are destined for farms already in disadvantaged areas with more difficult access to essential services and resources, hence not highly productive (Vigani and Dwyer, 2020). All in all, RDP payments are of voluntary nature, related to very specific investment projects, paving the way for distortive and subsidy-seeking farm practices (Khafagy and Vigani, 2022).

Zhu and Lansink (2010) assert that the negative effects of subsidies on farms' technical efficiency may relate to the demotivation of investing more resources in productivity-enhancing practices, as well as in more economical and environmentally efficient strategies. Likewise, Rizov, Pokrivcak and Ciaian (2013) explain how subsidies may resemble a distortion for farmers' production structure hence factor use, as they would simply look for receiving subsidies, even when this means investing in less-efficient and productive activities.

Kornai (1986) advised some time ago on the distortive role of subsidies and how they may lead to inefficient resource allocation as additional costs are not borne by the farm itself but by a different

agent – the public. This may bring the farmer towards riskier decisions he/she would make when directly incurring its costs, hence leading to moral hazard behaviours. Moreover, because subsidies may maintain inefficient farms in the sector, capital is shifted towards these less competitive agricultural units merely on a political basis, lowering the productivity of the whole system (Khafagy and Vigani, 2022).

Interestingly, when sub-samples according to the productivity level in each country are analyzed, discrepancies arise more often. Accordingly, it is important to stress that subsidies may not apply equally to all countries and farm typologies, feeding the idea of a more tailored and differentiated CAP. In line with Khafagy and Vigani (2022), we agree that a 'one-size fits all' CAP is far from being efficient and fair, and frequent measurements of subsidies' effectiveness are needed, with CAP having to be revised and steered according to different agricultural systems. Indeed, the results unveil that the extent to which policy instruments shape the farms' productivity, besides other factors, is significant and generally negative. However, different policy instruments serve different objectives – e.g., AES payments that are designed for enhancing the production of environmentally-related public goods or DDP aiming at supporting farmers' income – their (potential) side effects on the farm's productivity should also be considered, as the trade-off between objectives and priorities of the CAP. In particular, DDP decreases the efficient allocation of resources because, according to several analyses, low-productivity firms will keep operating instead of downsizing or exiting (Schiantarelli, 2008; Restuccia and Rogerson, 2013; Andrews and Cingano, 2014).

Nevertheless, how to efficiently design tailored subsidies according to the structure of the farms belonging to a specific agricultural sector is not, as well as which characteristics should be considered, is not an easy task. The present research hints that productivity levels could be a starting point, although we left to further research to define the method by which productivity shall be measured.

The results of the analyses allow for two main policy recommendations that could feed the debate on how to adjust the policies to cope with the recent turmoil observed in the cereal markets. The first refers to the possible use of coupled direct payments. If fostering cereal production becomes a policy priority, one option to consider is to increase the amount of direct payments coupled with cereal crops (i.e., CDP). This clearly will come at the cost of reducing the level of decoupled direct payments if cost of the overall policy should be maintained under control. Our analysis suggests that such a shift of resources from DDP to CDP could not harm the productivity of cereal farms in at least half of the considered countries, i.e., those where CDP is not found to have a negative effect on TFP. The second policy recommendation refers to the agro-environmental schemes (AES). Our analysis has confirmed that the support provided by AES exert a negative effect on the TFP of the cereal farms in almost all considered countries. Again, if increasing cereal production is considered important, it should be deemed to reconsider the extent and the design of AES to reduce the negative consequences these have on productivity. Clearly, this latter strategy should consider to ensure AES will continue to generate an adequate level of environmental benefits.

Finally, we would like to mention that the current analysis has some limitations, one of which is worth mentioning. This study does not explain the reasons for the differences encountered among countries. Further research and specific additional methods are needed to address this issue.

# Appendix

## A.1 Variables definition

For the definition of variables used we has followed the appendix 1 on Rizov, Pokrivcak and Ciaian, (2013),. In particular:

- *output value* $(Y)$ defined as the real value of total annual output (SE131)[10]
- $N$ t is total full-time equivalent labour input (SE011) measured in hours worked annually
- $K$ is capital is defined as the Total fixed assets (SE441) minus agricultural land value (in EU FADN dataset is ALNDAGR_CV_X)
- Estimate Land value we have followed the procedure of (Rizov et al., 2013) to include rental land. First, we determine the rental value of land, then, we evaluate the rate of return of land and finally, we discount the rental payments using the estimated rate of return as a discount rate.
    - The UAA hold is calculated
    
    $$Total\ UAA\ (SE025) - Rented\ UAA\ (SE030)$$
    
    - Value of fixed assets (SE441) per UAA hold is defined as
    
    $$\text{Fix\_K\_UAA\_hold } = \text{value of fixed assets /UAA hold}$$
    
    - Rent per UAA rented
    
    $$Rent_{UAA} = Rent\ paid \frac{SE375}{Rented} U.A.A. (SE030)$$
    
    - Rate of Return of Rent at farm level
    
    $$RoRR - Farm\ level = \frac{Rent_{UAA}}{Fix_{K_{UAA_{hold}}}}$$
    
    - Rate of Return of Rent for NUTS2 for FADN TF1 farm type using the median of $RoRR - Farm\ level$.
    - Regional aggregation through the use of the median rather than the estimated value per single company is necessary to avoid inconsistencies or company values that are too high or too low due to specific characteristics that cannot be repeated in other companies (for example, cases of land speculation): $RoRR - Aggregated$
    - Finally, we use $RoRR - Aggregated$ to discount the total rental payments (SE375), which gives us the capital value of land.
- $L$ Land value is given by the sum of the capital value of the land and agricultural land value (in FADN dataset is ALNDAGR_CV_X)

---

[10] FADN variable codes (FADN, 2014)

- $M$ is materials measured variable costs consisting of total annual specific costs (SE281) plus total annual farming overheads (SE336). SE281 represents current costs specific to production conversly SE 336 includes costs related to fixed capital (machinery and building costs).
- For CAP measure, we consider:
    - Total CAP subsidies $G$ is given by
    
    $$Total\ subsidies - excluding\ on\ investments\ (SE605)$$
    $$+ Subsidies\ on\ investments\ (SE406)$$
    
    - CDP Coupled Direct Payments
    - $CDP = Total\ subsidies\ on\ crops\ (SE610) +$
    $Total\ subsidies\ on\ livestock\ (SE615)$
    - $DDP = Direct\ Decoupled\ Payments\ (SE630)$
    - $RDPa$ = Rural Development Payments excluding investment
    
    $RDPa =$
    $Total\ support\ for\ rural\ development\ (SE624) - Subsidies\ on\ investments\ (SE406)$
    - $AES = Rural\ Development\ Payments\ for\ Agri-$
    $Environmental\ Schemes\ (SE621)$
    - $LFA =  Rural\ Development\ Payments\ for\ Less\ Favourable\ Areas\ (SE622)$
    - RDP Other Rural Development Payments excluding investment
    
    $$= RDPa - AES - LFA$$

Finally, in order to reduce the possible omitted price bias, we use the specific price indices, which allow us to have a proxy of the quantities of inputs and outputs (Zhu and Oude Lansink, 2008; Loecker, 2011; Rizov, Pokrivcak and Ciaian, 2013; Bond *et al.*, 2021).

## A.2 Minimum distance estimators

The reduced form of production function is the following:

$$y_{i,t} = \rho y_{i,t-1} + \beta_k k_{i,t} - \rho \beta_k k_{i,t-1} + \beta_n n_{i,t} - \rho \beta_n n_{i,t-1} + \beta_l l_{i,t} - \rho \beta_l l_{i,t-1}$$
$$+ \beta_m m_{i,t} - \rho \beta_m m_{i,t-1} + \beta_g g_{i,t} - \rho \beta_g g_{i,t-1} + (1-\rho)\eta_i \quad (8)$$
$$+ (\omega_{i,t} - \rho \omega_{i,t-1}) + (\varepsilon_{i,t} - \rho \varepsilon_{i,t-1})$$

To avoid excessive recourse to notations, without loss in clarity, we include the production factors in a generic factor $x$ and the term $(1-\rho)\eta_i + (\omega_{i,t} - \rho\omega_{i,t-1}) + (\varepsilon_{i,t} - \rho\varepsilon_{i,t-1}) = v_{i,t}$

$$y_{i,t} = \rho y_{i,t-1} + \beta_x x_{i,t} - \rho \beta_x x_{i,t-1} + v_{i,t} \quad (9)$$

SYS-GMM obtain the following model:

$$y_{i,t} = \pi_1 y_{i,t-1} + \pi_2 x_{i,t} + \pi_3 x_{i,t-1} + v_{i,t} \quad (10)$$

Consequently, the coefficients found should satisfy the mapping $\pi_1 = \rho$, $\pi_2 = \beta_x$, $\pi_3 = \rho\beta_x$

Given consistent estimates of the unrestricted coefficients in (9) the common factor restriction can be imposed using a minimum distance estimator following (Blundell et al., 1996; Chamberlain, 1982)

Letting $\pi = (\pi_1, \pi_2, \pi_3)$ and $\Theta = (\beta, \rho)'$, the restriction is $\pi(\Theta) = (\beta, -\beta\rho, \rho)'$

The restricted parameter estimates $\Theta$ are chosen to minimise the quadratic distance:

$$min\ [g(\hat{\pi}) - g(\pi(\Theta))]'\hat{\Omega}^{-1}[g(\hat{\pi}) - g(\pi(\Theta))] \qquad (11)$$

Where $\hat{\Omega} = (\partial g(\hat{\pi})/\partial \pi')\widehat{var}(\hat{\pi})(\partial g(\hat{\pi})/\partial \pi')'$ and $g(\pi) = (\pi_2, -\pi_3/\pi_1, \pi_1)'$ is chosen to make $g(\pi(\Theta)) = (\beta, \beta, \rho)'$ linear in $\beta$ and $\rho$

## A.3 Other Tables

**Table A. 1 - General Statistics in logarithm**

|  | GERMANY (DE) | | SPAIN (ES) | | FRANCE (FR) | | ITALY (IT) | | POLAND (PL) | | UNITED KINGDOM (UK) | |
|---|---|---|---|---|---|---|---|---|---|---|---|---|
| NR. OBS. | 22390 | | 20528 | | 19248 | | 29608 | | 31819 | | 5261 | |
| TOTAL OUTPUT | 12.04 | *(1.07)* | 10.69 | *(0.87)* | 12.03 | *(0.75)* | 10.66 | *(1.19)* | 10.61 | *(1.11)* | 12.57 | *(0.94)* |
| CAPITAL | 13.32 | *(1.21)* | 11.96 | *(1.09)* | 12.1 | *(1.02)* | 12.13 | *(1.65)* | 12.32 | *(0.97)* | 14.28 | *(1.32)* |
| LAND | 10.62 | *(2.54)* | 10.07 | *(2.42)* | 9.54 | *(1.55)* | 10.18 | *(2.71)* | 9.69 | *(2.53)* | 11.84 | *(2.93)* |
| LABOUR | 8.24 | *(0.81)* | 7.69 | *(0.57)* | 7.82 | *(0.56)* | 7.89 | *(0.69)* | 8.3 | *(0.60)* | 8.38 | *(0.96)* |
| MATERIAL | 11.62 | *(0.99)* | 10.2 | *(0.86)* | 11.61 | *(0.66)* | 9.94 | *(1.20)* | 10.09 | *(1.09)* | 12.16 | *(0.86)* |
| CAP | 10.56 | *(1.02)* | 9.47 | *(1.03)* | 10.48 | *(0.73)* | 9.07 | *(1.42)* | 9.4 | *(1.06)* | 10.91 | *(0.86)* |

*Note: Standard deviation in brackets*

*Source: authors' elaboration on FADN data*

**Table A. 2 – General Statistics: Values on average**

| Description | Average | Value on average | | | | | |
|---|---|---|---|---|---|---|---|
| | | Germany (DE) | Spain (ES) | France (FR) | Italy (IT) | Poland (PL) | United Kingdom (UK) |
| TOTAL OUTPUT | 212,360 | 1.56 | 0.31 | 1.04 | 0.45 | 0.42 | 2.22 |
| CAPITAL | 937,565 | 1.12 | 0.29 | 0.28 | 0.61 | 0.39 | 3.32 |
| LAND | 411,047 | 0.83 | 0.36 | 0.10 | 0.82 | 0.30 | 3.59 |
| LABOUR | 4,605 | 1.32 | 0.57 | 0.66 | 0.77 | 1.14 | 1.54 |
| MATERIAL | 128,768 | 1.60 | 0.31 | 1.07 | 0.36 | 0.43 | 2.23 |
| CAP SUBSIDIES | 44,052 | 1.64 | 0.47 | 1.02 | 0.52 | 0.53 | 1.82 |

*Source: authors' elaboration on FADN data*

**Table A. 3 – General Statistics: Indexes**

| Index | Germany (DE) | Spain (ES) | France (FR) | Italy (IT) | Poland (PL) | United Kingdom (UK) |
|---|---|---|---|---|---|---|
| **CAPITAL/LAND** | 3.07 | 1.83 | 6.17 | 1.70 | 2.97 | 2.11 |
| **LAND/LABOUR** | 56.00 | 56.05 | 14.10 | 94.76 | 23.34 | 208.50 |
| **CAPITAL /LABOUR** | 171.66 | 102.62 | 87.01 | 161.06 | 69.21 | 439.31 |
| **CAP/ TOTAL OUTPUT** | 0.22 | 0.32 | 0.20 | 0.24 | 0.26 | 0.17 |

**Table A. 4 - Variations of TFP between six countries**

| Year | Germany (DE) Mean | Delta | % | Cumulated (2009= 1.000) | Spain (ES) Mean | Delta | % | Cumulated (2009= 1.000) | France (FR) Mean | Delta | % | Cumulated (2009= 1.000) |
|---|---|---|---|---|---|---|---|---|---|---|---|---|
| 2009 | 1.549 | | | 1.000 | 1.422 | | | 1.000 | 0.878 | | | 1.000 |
| 2010 | 1.592 | 0.043 | 2.78% | 1.028 | 1.433 | 0.011 | 0.77% | 1.008 | 0.903 | 0.025 | 2.85% | 1.028 |
| 2011 | 1.589 | -0.003 | -0.19% | 0.998 | 1.432 | -0.001 | -0.07% | 0.999 | 0.907 | 0.004 | 0.44% | 1.004 |
| 2012 | 1.611 | 0.022 | 1.38% | 1.014 | 1.424 | -0.008 | -0.56% | 0.994 | 0.912 | 0.005 | 0.55% | 1.006 |
| 2013 | 1.599 | -0.012 | -0.74% | 0.993 | 1.422 | -0.002 | -0.14% | 0.999 | 0.898 | -0.014 | -1.54% | 0.985 |
| 2014 | 1.683 | 0.084 | 5.25% | 1.053 | 1.363 | -0.059 | -4.15% | 0.959 | 0.917 | 0.019 | 2.12% | 1.021 |
| 2015 | 1.556 | -0.127 | -7.55% | 0.925 | 1.387 | 0.024 | 1.76% | 1.018 | 0.901 | -0.016 | -1.74% | 0.983 |
| 2016 | 1.557 | 0.001 | 0.06% | 1.001 | 1.403 | 0.016 | 1.15% | 1.012 | 0.889 | -0.012 | -1.33% | 0.987 |
| 2017 | 1.554 | -0.003 | -0.19% | 0.998 | 1.394 | -0.009 | -0.64% | 0.994 | 0.9 | 0.011 | 1.24% | 1.012 |
| 2018 | 1.550 | -0.004 | -0.26% | 0.997 | 1.405 | 0.011 | 0.79% | 1.008 | 0.902 | 0.002 | 0.22% | 1.002 |

| Year | Italy (IT) Mean | Delta | % | Cumulated (2009= 1.000) | Poland (PL) Mean | Delta | % | Cumulated (2009= 1.000) | United Kingdom (UK) Mean | Delta | % | Cumulated (2009= 1.000) |
|---|---|---|---|---|---|---|---|---|---|---|---|---|
| 2009 | 1.068 | | | 1.000 | 1.053 | | | 1.000 | 0.807 | | | 1.000 |
| 2010 | 1.066 | -0.002 | -0.19% | 0.998 | 1.018 | -0.035 | -3.32% | 0.967 | 0.823 | 0.016 | 1.98% | 1.020 |
| 2011 | 1.065 | -0.001 | -0.09% | 0.999 | 1.005 | -0.013 | -1.28% | 0.987 | 0.826 | 0.003 | 0.36% | 1.004 |
| 2012 | 1.063 | -0.002 | -0.19% | 0.998 | 1.014 | 0.009 | 0.90% | 1.009 | 0.819 | -0.007 | -0.85% | 0.992 |
| 2013 | 1.060 | -0.003 | -0.28% | 0.997 | 1.006 | -0.008 | -0.79% | 0.992 | 0.816 | -0.003 | -0.37% | 0.996 |
| 2014 | 1.102 | 0.042 | 3.96% | 1.040 | 0.964 | -0.042 | -4.17% | 0.958 | 0.811 | -0.005 | -0.61% | 0.994 |
| 2015 | 1.069 | -0.033 | -2.99% | 0.970 | 0.993 | 0.029 | 3.01% | 1.030 | 0.819 | 0.008 | 0.99% | 1.010 |
| 2016 | 1.069 | 0.000 | 0.00% | 1.000 | 0.979 | -0.014 | -1.41% | 0.986 | 0.820 | 0.001 | 0.12% | 1.001 |
| 2017 | 1.066 | -0.003 | -0.28% | 0.997 | 0.991 | 0.012 | 1.23% | 1.012 | 0.820 | 0 | 0.00% | 1.000 |
| 2018 | 1.068 | 0.002 | 0.19% | 1.002 | 0.985 | -0.006 | -0.61% | 0.994 | 0.825 | 0.005 | 0.61% | 1.006 |

**Table A. 5 - TFP Estimation using Ackerberg, Caves, and Frazer (2015) Approach - (Step 2b)**

| Year | Germany (DE) Mean | St.d. | Spain (ES) Mean | St.d. | France (FR) Mean | St.d. | Italy (IT) Mean | St.d. | Poland (PL) Mean | St.d. | United Kingdom (UKI) Mean | St.d. |
|---|---|---|---|---|---|---|---|---|---|---|---|---|
| 2009 | 6.059 | *(0.657)* | 13.059 | *(1.189)* | 5.523 | *(0.542)* | 8.943 | *(1.273)* | 4.665 | *(1.376)* | 8.100 | *(0.651)* |
| 2010 | 5.871 | *(0.673)* | 12.686 | *(1.184)* | 5.328 | *(0.555)* | 8.843 | *(1.256)* | 3.992 | *(1.395)* | 7.916 | *(0.678)* |
| 2011 | 6.068 | *(0.664)* | 12.875 | *(1.153)* | 5.540 | *(0.546)* | 8.876 | *(1.250)* | 4.341 | *(1.337)* | 8.174 | *(0.663)* |
| 2012 | 6.085 | *(0.659)* | 13.024 | *(1.153)* | 5.630 | *(0.547)* | 8.925 | *(1.226)* | 4.368 | *(1.353)* | 8.186 | *(0.675)* |
| 2013 | 6.258 | *(0.670)* | 13.087 | *(1.126)* | 5.671 | *(0.607)* | 8.958 | *(1.238)* | 4.366 | *(1.428)* | 8.273 | *(0.686)* |
| 2014 | 6.135 | *(0.711)* | 13.060 | *(1.067)* | 5.488 | *(0.608)* | 8.872 | *(1.242)* | 4.213 | *(1.346)* | 8.116 | *(0.674)* |
| 2015 | 6.234 | *(0.708)* | 13.135 | *(1.038)* | 5.534 | *(0.549)* | 8.978 | *(1.306)* | 4.423 | *(1.316)* | 8.254 | *(0.624)* |
| 2016 | 6.194 | *(0.701)* | 13.254 | *(1.051)* | 5.568 | *(0.529)* | 9.125 | *(1.177)* | 4.256 | *(1.224)* | 8.264 | *(0.613)* |
| 2017 | 6.138 | *(0.696)* | 13.394 | *(1.085)* | 5.432 | *(0.525)* | 9.160 | *(1.178)* | 4.207 | *(1.255)* | 8.161 | *(0.643)* |
| 2018 | 6.143 | *(0.673)* | 13.392 | *(1.131)* | 5.559 | *(0.529)* | 9.119 | *(1.149)* | 4.301 | *(1.256)* | 8.170 | *(0.640)* |

*Standard deviation is indicated between the brackets in italics*    *Source: our elaboration on FADN data*

# Estimate of impact of CAP subsidies in farms with Low, Medium and High absolute levels of TFP by countries

Table A. 6 - Impacts of farm subsidies on TFP on Germany (DE): for Low, Medium and High absolute levels of TFP

|  | Germany (DE) | | |
|---|---|---|---|
|  | Low | Medium | High |
| $TFP_{t-1}$ | −0.039 | −0.004 | 0.033 |
|  | (0.039) | (0.007) | (0.066) |
| Economic Size (x1000) | 1.881* | 0.604*** | 0.045 |
|  | (1.041) | (0.155) | (0.120) |
| CDP (x1000) | −0.040 | −0.005 | 0.011 |
|  | (0.100) | (0.005) | (0.011) |
| DDP (x1000) | −0.009** | −0.008*** | −0.001 |
|  | (0.004) | (0.001) | (0.001) |
| AES (x1000) | −0.008* | −0.005*** | −0.006 |
|  | (0.005) | (0.001) | (0.006) |
| LFA (x1000) | 0.006 | −0.0001 | −0.030 |
|  | (0.048) | (0.003) | (0.041) |
| RDP Other (x1000) | −0.034 | −0.044*** | −0.121 |
|  | (0.084) | (0.006) | (0.266) |
| RDP inv (x1000) | −0.146 | −0.046*** | −0.106 |
|  | (0.090) | (0.006) | (0.266) |
| Sargan test (p.value) | 0.541 | 0.082 | 0.119 |
| m1 (p.value) | 0.000 | 0.002 | 0.000 |
| m2 (p.value) | 0.295 | 0.439 | 0.561 |
| Wald coeff. (p.value) | 0.000 | 0.000 | 0.001 |
| Wald time (p.value) | 0.000 | 0.000 | 0.000 |

**Table A. 7 - Impacts of farm subsidies on TFP on Spain (ES) for Low, Medium and High absolute levels of TFP**

|  | Spain (ES) | | |
|---|---|---|---|
|  | Low | Medium | High |
| $TFP_{t-1}$ | −0.053 | 0.007 | 0.020* |
|  | (0.068) | (0.043) | (0.012) |
| Economic Size (x1000) | 9.016*** | 1.144* | 1.999* |
|  | (2.089) | (0.590) | (1.111) |
| CDP (x1000) | −0.005 | −0.007** | −0.001 |
|  | (0.011) | (0.003) | (0.001) |
| DDP (x1000) | 0.002 | −0.001 | 0.002 |
|  | (0.008) | (0.002) | (0.001) |
| AES (x1000) | −0.016 | −0.006 | −0.001 |
|  | (0.015) | (0.004) | (0.002) |
| LFA (x1000) | −0.090** | −0.082*** | −0.018*** |
|  | (0.037) | (0.014) | (0.006) |
| RDP Other (x1000) | 0.017 | 0.368** | 0.004 |
|  | (0.096) | (0.179) | (0.003) |
| RDP inv (x1000) | 0.216 | 0.396** | 0.009** |
|  | (0.410) | (0.179) | (0.004) |
| Sargan test (p.value) | 0.782 | 0.080 | 0.029 |
| m1 (p.value) | 0.000 | 0.000 | 0.000 |
| m2 (p.value) | 0.758 | 0.658 | 0.391 |
| Wald coeff. (p.value) | 0.000 | 0.000 | 0.000 |
| Wald time (p.value) | 0.000 | 0.000 | 0.000 |

**Table A. 8 - Impacts of farm subsidies on TFP on France (FR) for Low, Medium and High absolute levels of TFP**

|  | France (FR) | | |
| --- | --- | --- | --- |
|  | Low | Medium | High |
| $TFP_{t-1}$ | −0.007 | −0.012 | −0.061** |
|  | (0.049) | (0.030) | (0.030) |
| Economic Size (x1000) | 0.278 | 0.088*** | 0.088*** |
|  | (0.485) | (0.033) | (0.010) |
| CDP (x1000) | 0.000 | −0.002*** | −0.001*** |
|  | (0.005) | (0.000) | (0.000) |
| DDP (x1000) | −0.003** | −0.0016*** | 0.0001 |
|  | (0.001) | (0.000) | (0.000) |
| AES (x1000) | −0.001 | −0.001* | −0.001*** |
|  | (0.001) | (0.000) | (0.000) |
| LFA (x1000) | −0.018 | −0.002*** | −0.003*** |
|  | (0.011) | (0.001) | (0.001) |
| RDP Other (x1000) | 0.006 | −0.000 | −0.002*** |
|  | (0.008) | (0.000) | (0.000) |
| RDP inv (x1000) | 0.013 | −0.001** | −0.003*** |
|  | (0.010) | (0.000) | (0.001) |
| Sargan test (p.value) | 0.470 | 0.099 | 0.656 |
| m1 (p.value) | 0.000 | 0.000 | 0.000 |
| m2 (p.value) | 0.674 | 0.482 | 0.322 |
| Wald coeff. (p.value) | 0.027 | 0.000 | 0.000 |
| Wald time (p.value) | 0.000 | 0.000 | 0.000 |

**Table A. 9 - Impacts of farm subsidies on TFP on Italy (IT) for Low, Medium and High absolute levels of TFP**

|  | Italy (IT) | | |
|---|---|---|---|
|  | Low | Medium | High |
| $TFP_{t-1}$ | 0.037 | 0.003 | 0.140*** |
|  | (0.042) | (0.060) | (0.038) |
| Economic Size (x1000) | 0.430** | 0.550 | 0.050 |
|  | (0.213) | (0.563) | (0.069) |
| CDP (x1000) | −0.066 | −0.008 | −0.001 |
|  | (0.053) | (0.020) | (0.001) |
| DDP (x1000) | −0.007*** | 0.0004 | −0.001*** |
|  | (0.002) | (0.001) | (0.000) |
| AES (x1000) | −0.003*** | 0.00002 | −0.0001 |
|  | (0.000) | (0.004) | (0.000) |
| LFA (x1000) | 0.006 | −0.011 | −0.0004 |
|  | (0.004) | (0.009) | (0.001) |
| RDP Other (x1000) | −0.010*** | −0.002 | −0.001* |
|  | (0.002) | (0.007) | (0.001) |
| RDP inv (x1000) | −0.010*** | −0.005 | −0.002** |
|  | (0.002) | (0.015) | (0.001) |
| Sargan test (p.value) | 0.302 | 0.715 | 0.110 |
| m1 (p.value) | 0.000 | 0.000 | 0.000 |
| m2 (p.value) | 0.490 | 0.572 | 0.492 |
| Wald coeff. (p.value) | 0.000 | 0.079 | 0.000 |
| Wald time (p.value) | 0.000 | 0.000 | 0.000 |

**Table A. 10 - Impacts of farm subsidies on TFP on Poland (PL) for Low, Medium and High absolute levels of TFP**

|  | Poland (PL) | | |
|---|---|---|---|
|  | Low | Medium | High |
|  | (1) | (2) | (3) |
| $TFP_{t-1}$ | 0.059 | 0.007 | 0.109*** |
|  | (0.049) | (0.049) | (0.028) |
| Economic Size (x1000) | 3.088*** | −0.047 | 0.488*** |
|  | (0.708) | (0.310) | (0.174) |
| CDP (x1000) | −0.005 | −0.002 | 0.001 |
|  | (0.009) | (0.003) | (0.001) |
| DDP (x1000) | −0.016*** | 0.004 | −0.002*** |
|  | (0.006) | (0.003) | (0.001) |
| AES (x1000) | 0.003 | −0.001 | −0.008*** |
|  | (0.010) | (0.006) | (0.002) |
| LFA (x1000) | −0.027 | −0.089*** | −0.003 |
|  | (0.045) | (0.023) | (0.002) |
| RDP Other (x1000) | −0.062 | 0.069 | 0.014 |
|  | (0.060) | (0.078) | (0.016) |
| RDP inv (x1000) | −0.069 | 0.067 | 0.013 |
|  | (0.061) | (0.078) | (0.016) |
| Sargan test (p.value) | 0.392 | 0.120 | 0.177 |
| m1 (p.value) | 0.000 | 0.000 | 0.000 |
| m2 (p.value) | 0.707 | 0.624 | 0.417 |
| Wald coeff. (p.value) | 0.001 | 0.005 | 0.000 |
| Wald time (p.value) | 0.000 | 0.000 | 0.000 |

**Table A. 11 - Impacts of farm subsidies on TFP on the United Kingdom (UK) for Low, Medium and High absolute levels of TFP[11]**

|  | United Kingdom (UK) | | |
|---|---|---|---|
|  | Low | Medium | High |
| $TFP_{t-1}$ | −0.014 | 0.111 | 0.199** |
|  | (0.119) | (0.089) | (0.102) |
| Economic Size (x1000) | −0.112 | −0.043** | −0.033** |
|  | (0.208) | (0.019) | (0.016) |
| CDP (x1000) | −0.008 | 0.001 | −0.005 |
|  | (0.005) | (0.001) | (0.004) |
| DDP (x1000) | 0.001 | −0.000 | 0.000 |
|  | (0.001) | (0.000) | (0.000) |
| AES (x1000) | −0.000 | 0.001 | 0.000 |
|  | (0.001) | (0.001) | (0.000) |
| RDPa Other (x1000) | 0.000 | −0.000 | −0.003 |
|  | (0.001) | (0.001) | (0.009) |
| RDP Inv (x1000) | 0.000 | −0.000 | −0.003 |
|  | (0.001) | (0.000) | (0.008) |
| Sargan test (p.value) | 0.051 | 0.955 | 0.053 |
| m1 (p.value) | 0.013 | 0.000 | 0.000 |
| m2 (p.value) | 0.194 | 0.099 | 0.979 |
| Wald coeff. (p.value) | 0.563 | 0.012 | 0.049 |
| Wald time (p.value) | 0.000 | 0.000 | 0.000 |

*Standard errors are indicated between the brackets in italics*     *Source: our elaboration on FADN data*

---

[11] In the United Kingdom, the farmers that participate in the RDPa LFA scheme in the sample are very low. This does not allow System- GMM estimator to reach solutions.